\documentclass[journal=jacsat,manuscript=article]{achemso}

\usepackage[version=3]{mhchem} 
\usepackage{gensymb}
\usepackage[]{textcomp}
\usepackage{amsmath}
\usepackage{amstext}
\usepackage{comment}
\usepackage{url}

\newcommand{\IA}{\AA$^{-1}$}

\author{Howie Joress}
\affiliation[NIST]{Materials Measurement Science Division, National Institute of 
 Standards and Technology, Gaithersburg, MD 20899}
 \email{howie.joress@nist.gov}
\author{Brian L. DeCost}
\affiliation[NIST]{Materials Measurement Science Division, National Institute of 
 Standards and Technology, Gaithersburg, MD 20899}
 \author{Suchismita Sarker}
\affiliation[SLAC]{Stanford Synchrotron Radiation Lightsource, SLAC National Accelerator Laboratory, Menlo Park, CA 94025, USA}
\author{Trevor M. Braun}
\affiliation[NIST]{Materials Science and Engineering Division, National Institute of 
 Standards and Technology, Gaithersburg, MD 20899}
\author{Sidra Jilani}
\affiliation{School of Materials Science and Engineering, University of New South Wales, Sydney, New South Wales 2052, Australia}
\author{Ryan Smith}
\affiliation[NIST]{Materials Measurement Science Division, National Institute of 
 Standards and Technology, Gaithersburg, MD 20899}
\author{Logan Ward}
\affiliation{Department of Materials and Engineering, Northwestern University, Evanston, Il 60208}
\author{Kevin J. Laws}
\affiliation[UNSW]{School of Materials Science and Engineering, University of New South Wales, Sydney, New South Wales 2052, Australia}
\author{Apurva Mehta}
\affiliation[SLAC]{Stanford Synchrotron Radiation Lightsource, SLAC National Accelerator Laboratory, Menlo Park, CA 94025, USA}
\author{Jason R. Hattrick-Simpers}
\affiliation[NIST]{Materials Measurement Science Division, National Institute of 
 Standards and Technology, Gaithersburg, MD 20899}

\title[]
  {A high-throughput structural and electrochemical study of  metallic glass formation in Ni-Ti-Al}

\abbreviations{}

\begin{document}

\begin{tocentry}

\includegraphics[]{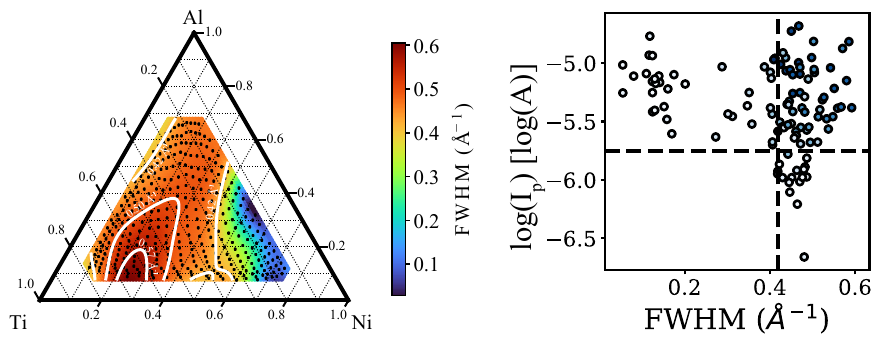}




 \end{tocentry}

\begin{abstract}
Based on a set of machine learning predictions of glass formation in the Ni-Ti-Al system, we have undertaken a high-throughput experimental study of that system.  We utilized rapid synthesis followed by high-throughput structural and electrochemical characterization.  Using this dual-modality approach, we are able to better classify the amorphous portion of the library, which we found to be the portion with a full-width-half-maximum (FWHM) of $\>0.42$ \IA{} for the first sharp x-ray diffraction peak.   We demonstrate that the FWHM and corrosion resistance are correlated but that, while chemistry still plays a role, a large FWHM is necessary for the best corrosion resistance.
\end{abstract}
\section{Introduction}
Metallic glasses are of great technological interest because they have been demonstrated to show superior properties, including high hardness and corrosion resistance, in comparison to their crystalline counterparts\cite{ashby,khan1,scully}.  To date, only around 6000 compositions have been explored and logged in tabulated formats\cite{kawazoe1997nonequilibrium,ren2018accelerated}. Therefore, there remain vast unexplored regions of composition space where metallic glasses with exciting properties may be found, as many as 3 million by some estimates\cite{li2017many}.    Unfortunately, it can be challenging to identify which alloys will have sufficient glass forming ability (GFA).  Due to the multiple length and time scales involved in solidification, accurate first-principles calculations of GFA have proven difficult\cite{perim2016spectral}.    In lieu of these models, physio-chemical based heuristic models have been developed.  These include the well-known Turnbull model based on deep eutectics\cite{marcus1976correlation} as well as more recent models including the efficient packing model \cite{laws2015predictive} and the model by \citet{yang2012prediction} based on atomic size variation and the ratio of the entropy to enthalpy of mixing.  These models provide some guidance in searches for new metallic glasses, but as of yet they have not been demonstrated to be global predictors of alloys with high GFA.  Recently, we published a machine learning (ML) based model for prediction of GFA that was iteratively coupled with high throughput synthesis and characterization\cite{ren2018accelerated}.  This methodology demonstrated a significant increase in the rate of discovery of new metallic glass (MG) systems, and after several iterations, the predictions were determined to be reliable enough to pursue even when they violate the above heuristics.  

High-throughput studies of MGs are complicated  due to the lack of methods for differentiating amorphous from crystalline samples.  This is particularly true in thin films where calorimetric techniques are difficult\cite{mccluskey2011glass}.  A common method in literature, applied to  both thin film and bulk samples is to measure the full width at half the maximum (FWHM) of the first sharp diffraction peak (FSDP) in the x-ray diffraction from the sample.  There is, however, no consensus on what values for this FWHM constitutes a MG.  Further it likely varies between systems.  Many papers simply give a subjective description of the diffraction data.  \citet{ma2009power} generalizes that MGs have a FWHM of between 0.4 \IA{} and 0.5 \IA{}\bibnote{\AA$=10^{-10} m$}. Ref. \citenum{ren2018accelerated} uses the FWHM for amorphous \ce{SiO2}, 0.57 \IA{}, as the threshold, a more conservative value than is typically applied in the literature. 

In this paper, we explore an alloy system, Ni-Ti-Al.  This elemental system was selected for study because the regions our machine learning model predicted to have high GFA differed greatly between the predictions for the bulk versus the stacked thin-film model.  Further, as will be reported below, the model predicted a large range of high GFA that extended far beyond the deep eutectic region located near the \ce{Ti3Ni} portion of the phase diagram\cite{budberg1992aluminium}. The regions with high potential GFA included large swaths of the Ti-Al and Ni-Al binaries. This could be a technologically important material systems as the presence of Al is likely to imbue the MG with the ability to resist corrosion in saline environments. For instance, this system could be of interest for coating of Nitinol (Ni-Ti alloys) stents, where dissolution and absorption of Nickel is potentially toxic and/or carcinogenic in nature\cite{thierry2000effect}.  Based on these interesting ML predictions we proceeded with a traditional high-throughput study of this system.

Using combinatorial thin-film synthesis we produced, in parallel, a range of alloys covering a large, central portion of the Ni-Ti-Al ternary system.  We then rapidly characterized the resulting library with synchrotron x-ray diffraction and electrochemical mapping.  We found that there is a region of glass formation that extends beyond that previously demonstrated in the literature.  
Through comparisons of the literature values used in this system to label alloys amorphous, the correlation of corrosion measurements with diffraction measurements, and consideration of the packing efficiency and electronic state of the alloys we determined that a FWHM of 0.42 \IA{} for the FSDP is a good threshold for declaring a given alloy amorphous in the NiTiAl system.  
We further observe that the deepest glass forming region, here determined by the region with the largest FWHM, was highly correlated with the bulk ML model, while relaxing the FWHM criteria resulted in a larger glass forming region consistent with the thin film ML model. 

\section{Experimental}
A process flow diagram showing the various parts of the study, their interaction, and the use of ML is included in the supplemental material.
\subsection{Machine Learning Model}
This elemental system was selected based upon the ML model described in Ref. \citenum{ren2018accelerated}.  In short the model is a random forest model\cite{breiman2001random} using the Magpie feature set as descriptors\cite{ward2016general}.  The probability of glass forming ability ($P_{GFA}$) was predicted using the model as trained at the end of that work (and therefore including all the data published within along with data from Ref. \citenum{kawazoe1997nonequilibrium}).  Two types of predictions were made:  The first uses only bulk MGs from literature.  The second is a stacked model, optimized for predicting GFA in thin films, trained using sputtered MG alloys with bulk model predictions as an input.

\subsection{Library Synthesis}
Thin-film combinatorial libraries were synthesized by co-sputtering a continuous composition spread from three elemental sources onto a 76 mm (3 inch) wafer.  The geometry of the sputter guns was such that the center of depositions for the three sources were off center from the wafer and the deposition rate for each source varied as a function of position across the wafer, and thereby a continuous composition spread is formed.  In this case we sputtered from metallic targets of Al, Ni, and Ti, each with purities of better than 99.99 \% with a power of 10.8 W/cm$^2$, 1.07 W/cm$^2$, and 4.28 W/cm$^2$ respectively.  Prior to sputtering, the base pressure of the chamber was better than 1.33×10$^{-4}$ Pa and sputtering was done under a Ar ($>$99.99 wt\% purity)background of 0.67 Pa.  The composition of the film was measured by scanning x-ray fluorescence (XRF) using a Bruker Tornado and analyzed by Crossroads XRS-MTFFP software \bibnote{Certain commercial equipment, instruments, software, or materials  are identified in this paper to foster understanding. Such identification does not imply recommendation or endorsement by the National Institute of Standards and Technology, nor does it imply that the materials or equipment identified are necessarily the best available for the purpose.}.  The film was mapped with 177 points with approximately a 0.02 mm spot-size \cite{hattrick2019inter}.  We interpolated these measurements across the wafer using a Gaussian process. 

\subsection{X-ray characterization}
The film was mapped using x-ray diffraction at the Stanford Synchrotron Radiation Light Source beamline 1-5\cite{gregoire2014high}.  Diffraction patterns for   362 points on the wafer were collected.
The details of the measurement and the code used to extract the full width half maximum (FWHM) of the first sharp diffraction peak (FSDP) are described in Refs. \citenum{ren2018accelerated,ren2017fly} and the source can be found at \url{https://github.com/fang-ren/Discover_MG_CoVZr}.   Where needed, a gaussian process was fit to the FWHM values, to allow for correlation with the electrochemical data.

\subsection{Electrochemical characterization}
\begin{figure}
    \centering
    \includegraphics[width=8.85cm]{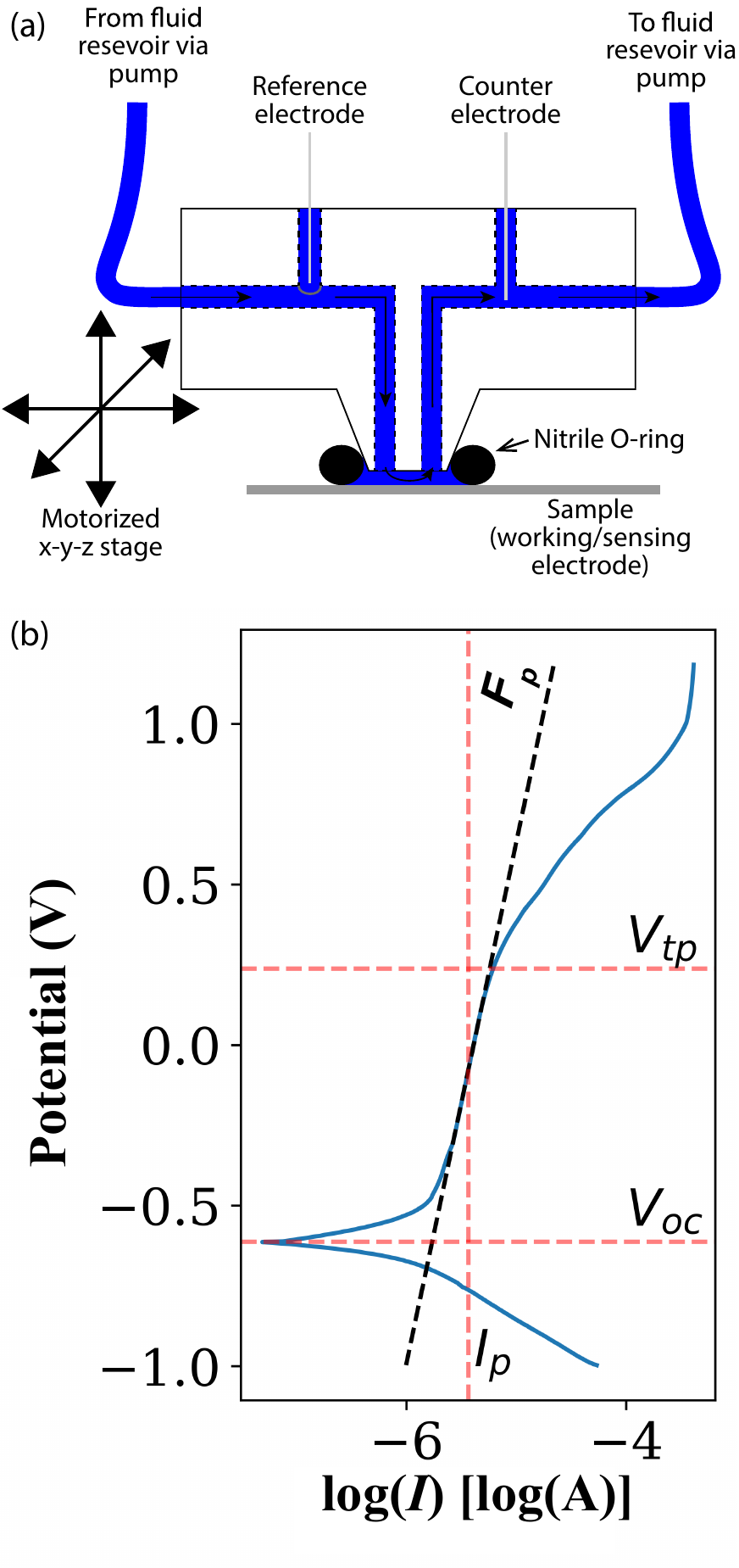}
    \caption{(a)Schematic showing the scan head of the SDC.  The electrolyte flows through the scan head passing by the reference and counter electrodes as illustrated.  The scan head is mounted on a motorized Z stage to enable contact of the scan head with the sample.  The sample is mounted underneath the scan head on a motorized x-y stage to allow for rastering of the scan head over the sample.  (b) a representative polarization curve showing extracted metrics.}
    \label{fig:SDC}
\end{figure}
For electrochemical characterization we used a scanning droplet cell (SDC), a millimeter scale electrochemical cell optimized for localized electrochemical measurements similar to those described in Ref. \citenum{woodhouse2009combinatorial,klemm2011high,klemm2011combinatorial,gregoire2013scanning}.  The SDC itself comprises a peristaltic pump, a fluid reservoir, and a scan head.  The scan head, shown schematically in Fig. \ref{fig:SDC} is a polytetrafluoroethylene block with a set of ports that can be sealed against the thin-film library with a nitrile o-ring.  The pump pushes the electrolyte from the fluid reservoir through the scan head into an isolated volume between the cell and the substrate and back to the reservoir.  As the fluid enters and exits this volume it passes by a counter electrode (Pt wire) and reference electrode (Ag/AgCl in a saturated KCl solution) respectively.  The metallic film acts as the working/sensing electrode.   The contact area for the electrolyte on the film is a circle with a 3.75 mm diameter and the measurements were taken on a 9 mm pitch checkerboard across the wafer.  Corrosion measurements were performed by acquiring polarization curves in 1.1 mol/L NaCl (brine) solution.  To ensure the corrosion measurements were representative of the base metal, the bias was first swept negative from the open circuit potential to clean the surface, then swept positive to corrode the metal.  Sweeping of the potential was done at a rate of 0.075 V/s and the total scan range was -1 V to 1.2 V. In total we measured 109 points across the wafer, averaging 4.5 minutes of measurement time per point.

The features of interest from the polarization curve were determined to be the the passivation current (I$_{\text{p}}$) and the flatness of the passivation plateau transpassive potential. The open circuit potential $V_{OC}$ and transpassive potential ($V_{\text{tp}}$) were also tabulated and are provided in the supplemental data.  We obtain these features from each polarization curve by an iterative fitting procedure, using the log absolute current as a function of potential curve, as illustrated in Fig. \ref{fig:SDC}(b).  The code can also be found in the SI.
First, we fit a Laplace (i.e. a double exponential) peak model to the polarization curve to identify the open circuit potential $V_{oc}$ (shown here just below -0.5 V).
This model uses a fifth-order polynomial background to ensure a good fit to the peak location and width.
We identify the onset of the passivation plateau as the potential corresponding to the 99th percentile of the Laplace peak model.
Starting from this onset value, we fit a series of robust linear regression models to the rest of the polarization curve, constraining the model to pass through the onset point.
We select the potential range for fitting these models by minimizing the mean squared error on the entire polarization curve.
The passivation plateau flatness, $F_p$ is simply the slope of this linear model.
Using this final robust linear model, we identify the transpassive potential $V_{\text{tp}}$ by performing automatic threshold selection on the model residuals (using the triangle method \cite{zack1977automatic}).
Finally, we report the passivation current I$_{\text{p}}$ as the median log current of the passivation plateau. We note that the geometry of this electrochemical cell creates non-uniformities in flow leading to uneven corrosion of the wetted portion of the film.  To this end, values should be taken as semi-quantitative; relative values are informative for  comparison but the absolute value of the corrosion metrics have not been verified.

\section{Results and discussion}

\subsection{Machine Learning Models and Literature Data}
\begin{figure}
    \centering
    \includegraphics[width=8.85cm]{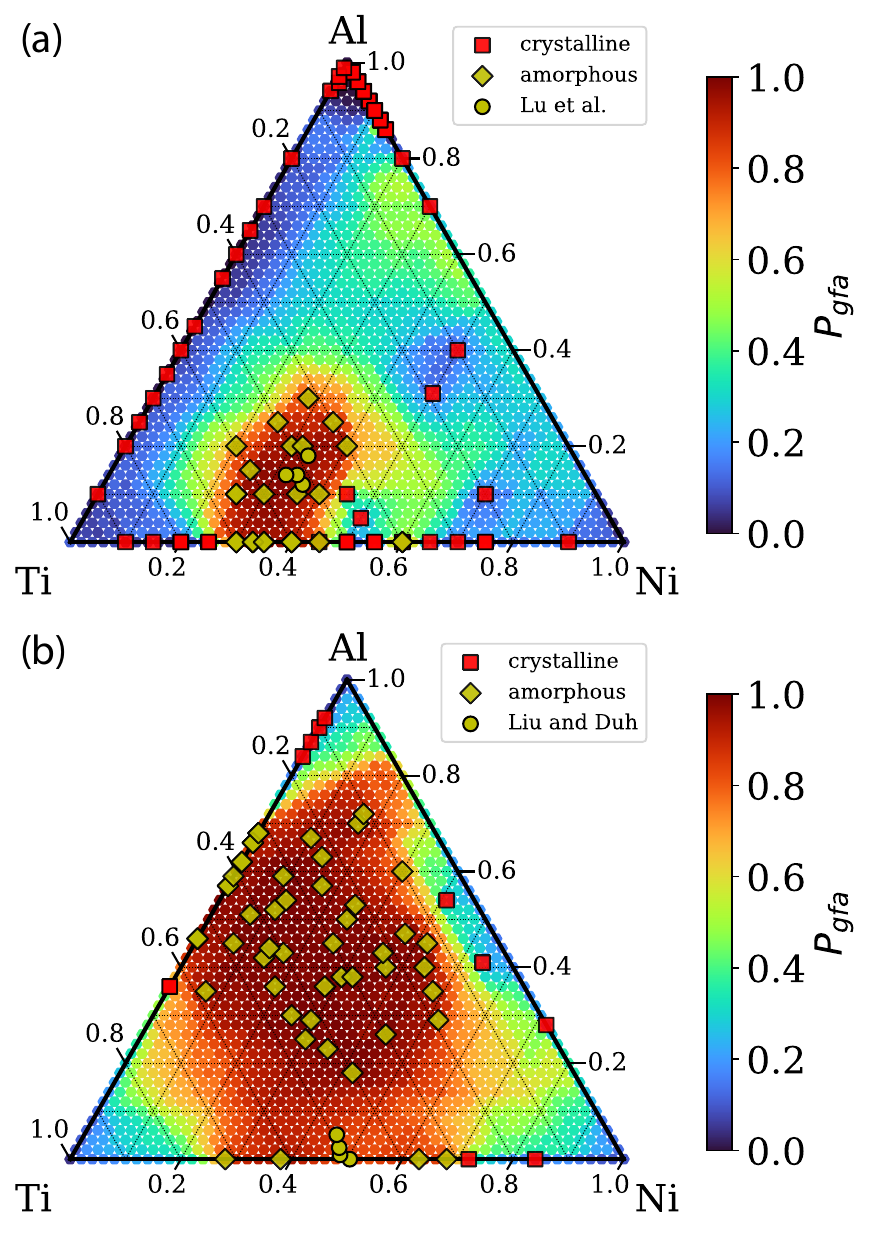}
    \caption{ML model predictions of MG formation for the Ni-Ti-Al system, displayed as a probability ($P_{GFA}$).  (a) shows the bulk model and (b) shows the stacked model for thin-films.  Amorphous and crystalline data contained in the training set are plotted on the models along with the amorphous data points from Ref. \citenum{lu2009optimal} and \citenum{liu2007effect} (not used in ML model). }
    \label{fig:ML}
\end{figure}
Fig. \ref{fig:ML} shows the ML predictions of the Ni-Ti-Al system along with previous data from the literature.  It is notable that  a large portion of the phase diagram between the Ni-Ti binary and alloys with less than 20 at\% Al have not been explored in thin-films. Importantly, this region contains alloys that are closest in Ni/Ti ratio to the ideal shape memory alloy composition of \ce{Ni30Ti55Cu15}\citet{cui2006combinatorial}.  MGs in this range would be likely to be chemically compatible and exhibit limited Ni/Ti diffusion with ideal shape memory alloy stents.

The bulk model, shown in Fig. \ref{fig:ML}(a), predicted a limited region of glass formability, with only the Ni-Ti binary region showing significant predicted GFA.  This region is highly correlated with the training data as well as the literature data from Ref. \citenum{lu2009optimal} and is located adjacent to the deep \ce{Ti3Ni} eutectic\cite{budberg1992aluminium}.  The stacked thin-film model, shown in Fig. \ref{fig:ML}(b) shows a much larger region of high predicted GFA, again well correlated with the training data with some interpolation between points.  It should be noted that these data, from Ref. \citenum{akiyama_habazaki_kawashima_asami_hashimoto_1993}, reported Al-Ti binaries with between $\approx$ 46 and 69 At\% Al to be amorphous, although the FWHM of the peak shown in their manuscript would not have resulted in its classification as a MG using our previous criteria. Moreover, the atomic radii of Ti and Al are quite close to one another and therefore these alloys could be considered topologically as a binary system, and therefore, ordinarily there would be little reason to expect such a large glass forming region. 


\subsection{Observed FWHM Values and Comparison to Literature}

\begin{figure}
    \centering
    \includegraphics[width=8.5cm]{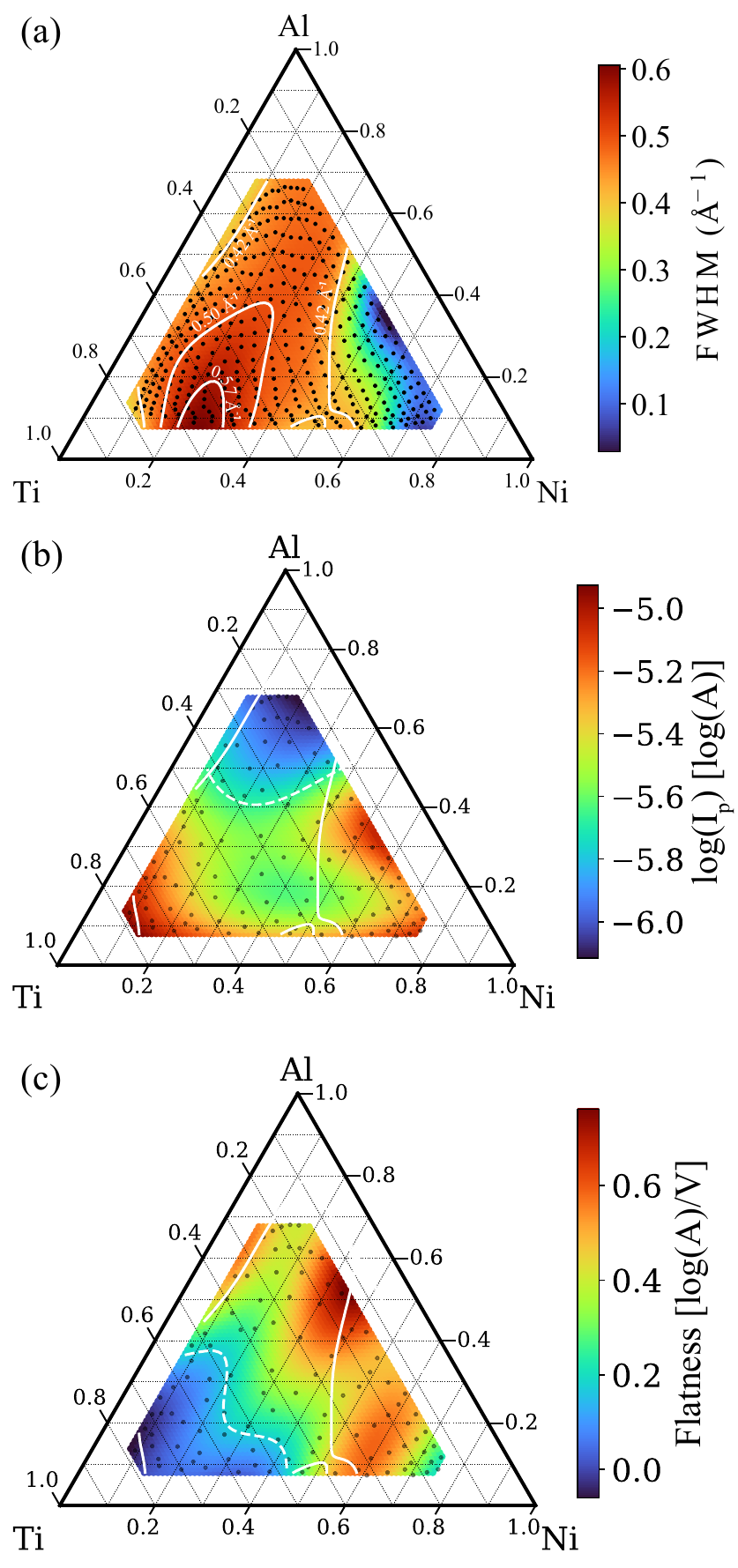}
    \caption{(a)Ternary plot showing the FWHM of the first bright diffraction peak as a function of chemistry.  Interpolated values, using a  Gaussian process, are plotted under points showing locations of XRD measurements.  Isolines for FWHM corresponding to 0.57 \IA{}, 0.5 \IA{}, and 0.42 \IA{} are included.  (b) and (c) show the results from the electrochemical cell, similarly plotted.  (b) shows the $\log(I_p)$.  Both are annotated with the bound of the glass region (solid white line) and optimal corrosion resistance (dotted white line).}
    \label{fig:results}
\end{figure}
Fig.\ref{fig:results}(a) shows the FWHM of the FSDP as measured by high-throughput x-ray diffraction from our sputtered film as a function of composition.  As we have previously reported, the sputtered films exhibit a FWHM that varies dramatically across the library from $\approx$ 0.1 \IA to $\approx$ 0.6 \IA.  The widest peak is seen in the region adjacent to the \ce{Ti3Ni} composition predicted by our bulk model to be glass forming and which was previously reported in literature.\cite{lu2009optimal,liu2007effect}.  Additionally, there is a valley in the FWHM extending from this region to \ce{Al3Ni}, a known intermetallic\cite{nash1991ni}, passing along the \ce{TiNi_{0.5}Al_{1.5}} intermetallic\cite{zeng1999ternary}.  The general trend FWHM is roughly consistent with the work of \citeauthor{akiyama_habazaki_kawashima_asami_hashimoto_1993}, where the inclusion of Al results in a sharpening of the FSDP. This observed extended range is similar to the thin film predictions, with the exception that it doesn't reach the Ti-Al binary edge. We previously posited that, in sputtered MG films, glass formation was highly correlated with the structural frustration associated with the quenched atoms being unable to order into their intermetallic structure, which is consistent with the observed variability in the FWHM\cite{ren2018accelerated}.

Despite the widths of XRD peaks being given as evidence for an amorphous structure, as mentioned above, there is no precise structural definition of amorphous alloys, particularly based on a measure of the FWHM of the FSDP.  In our previous work, in the Co-Zr-X (x=V, Fe, and Ti), we used 0.57 \IA, based on amorphous \ce{SiO2}, as the minimum FWHM of the FSDP for a glass.  This value is very conservative, particularly in comparison to other published values.  For instance, for the Ni-Ti-Al system there is a great variety of FWHM values that are given as evidence of an amorphous alloy.  Using digital pixel-to-scale correlation, we were able to extract the FWHM of the FSDP from XRD patterns published in literature. Ref. \citenum{lu2009optimal} correlates XRD with a $\approx$ 0.4 \IA{} FWHM of the FSDP with an amorphous phase for bulk alloys.  Ref. \citenum{liu2007effect} uses and even narrower $\approx$ 0.33 \IA{} peak for classifying a thin film alloy as glassy.  In the case of Ref. \citenum{akiyama_habazaki_kawashima_asami_hashimoto_1993}, which forms the basis of the MGs in our training set, the minimum FWHM used to determine that a material was glassy was $\approx$ 0.28 \IA{} for Al66Ti34 (The largest FWHM for published XRD data was $\approx$ 0.38 \IA{}). The portion of composition space that is amorphous given a range of FWHM cutoffs varies dramatically, as seen in Fig. \ref{fig:results}(a). For the most conservative estimates of $\approx$ 0.57 \IA{} FWHM, only the region predicted by the bulk model can be considered to be amorphous. If, however, the criteria is systematically relaxed to be more consistent with the smaller literature values, then the range of observed glasses extends along the previously described valley and includes all measured alloys between 10 at.\% - 50 at.\% percent Ni. This is potentially much more consistent with the thin film model.


\subsection{Corrosion mapping}


Figs. \ref{fig:results}(b)-(c) show two corrosion metrics extracted from the polarization curves measured using the scanning droplet cell: passivation current ($I_p$) and flatness of the passivation plateau ($F_{p}$) and   (the other metrics, $V_{tp}$ and $V_{oc}$, are noisy so further analysis is difficult; they are included in the SI).

The $V_{OC}$ during the original anodic sweep is a measure of the mixed potential of the clean metal surface and the various reactants present in the electrolyte. As a function of composition, $V_{OC}$ in this system largely follows a rule of mixtures,  is contained in the supplemental materials. 
The general trend is that the Nickel part of the phase diagram is the most noble (less negative standard reduction potential), while Ti and Al are more reactive.  The $I_{p}$ is taken to be the average of the log corrosion current in the passivating regime and is an analog for corrosion rate of the passivated surface.  Here, the $I_{P}$ shows good correlation with the FWHM, a clear valley in current density is visible in the range of alloys that exhibit the highest FWHM. In addition, the lowest values of $I_{p}$ are observed in the regions with the highest Al concentration, this is sensible as Al is known to form passive films. The $F_{p}$ of the plateau region is an estimate of the "stability" of the oxide scale growth rate to small perturbations during operation, with a small value being more stable.   In this case there is also good correlation between $F_{p}$ and the FWHM. The chemistry of the alloy is important here as well, with the Ti-rich samples exhibiting the lowest $F_p$. 

\citeauthor{akiyama_habazaki_kawashima_asami_hashimoto_1993} investigated the corrosion of Ni-Ti-Al alloys in 1 mol/L HCl and it is illustrative to compare the compositional trends between their work and the present work. The $I_{P}$ mapping is most comparable to corrosion rate (reported in mm/y). In the case of immersion in aerated HCl, increasing the Al content resulted in an order of magnitude increase in the corrosion rate from 0.1 mm/year to 2.0 mm/year.  The corrosion rate was relatively insensitive to the ratio of Ni to Ti, although from the published diffraction data it was correlated to the diffraction FWHM. Conversely, in the case of 1 mol/L NaCl, we see that Al additions are responsible for an order of magnitude decrease in the corrosion current and the alloys containing the most Ti and Ni (and also the largest FWHM) corrode the quickest. From the Pourbaix diagram, neither Ni nor Al would  be expected to be passivating in the presence of 1 mol/L HCl, although Ni could act to ennoble the open circuit potential, thus reducing the corrosion rate. Near neutral pH none of the three elements will actively dissolve and thus the corrosion rate is determined by passivity of the oxide or hydroxide. A detailed study of the phase and dynamics of the formed oxides will be reported elsewhere.

\subsection{Correlating corrosion with glass formation}

Metallic glasses will generally exhibit better resistance to corrosion than their crystalline counterparts, as high energy crystalline defects (eg. grain boundaries and lattice vacancies) are unavailable. Therefore, the electrochemical corrosion data provides a second method of evaluating the glassy nature of our films. 
\begin{figure}
    \centering
    \includegraphics[width=8.5cm]{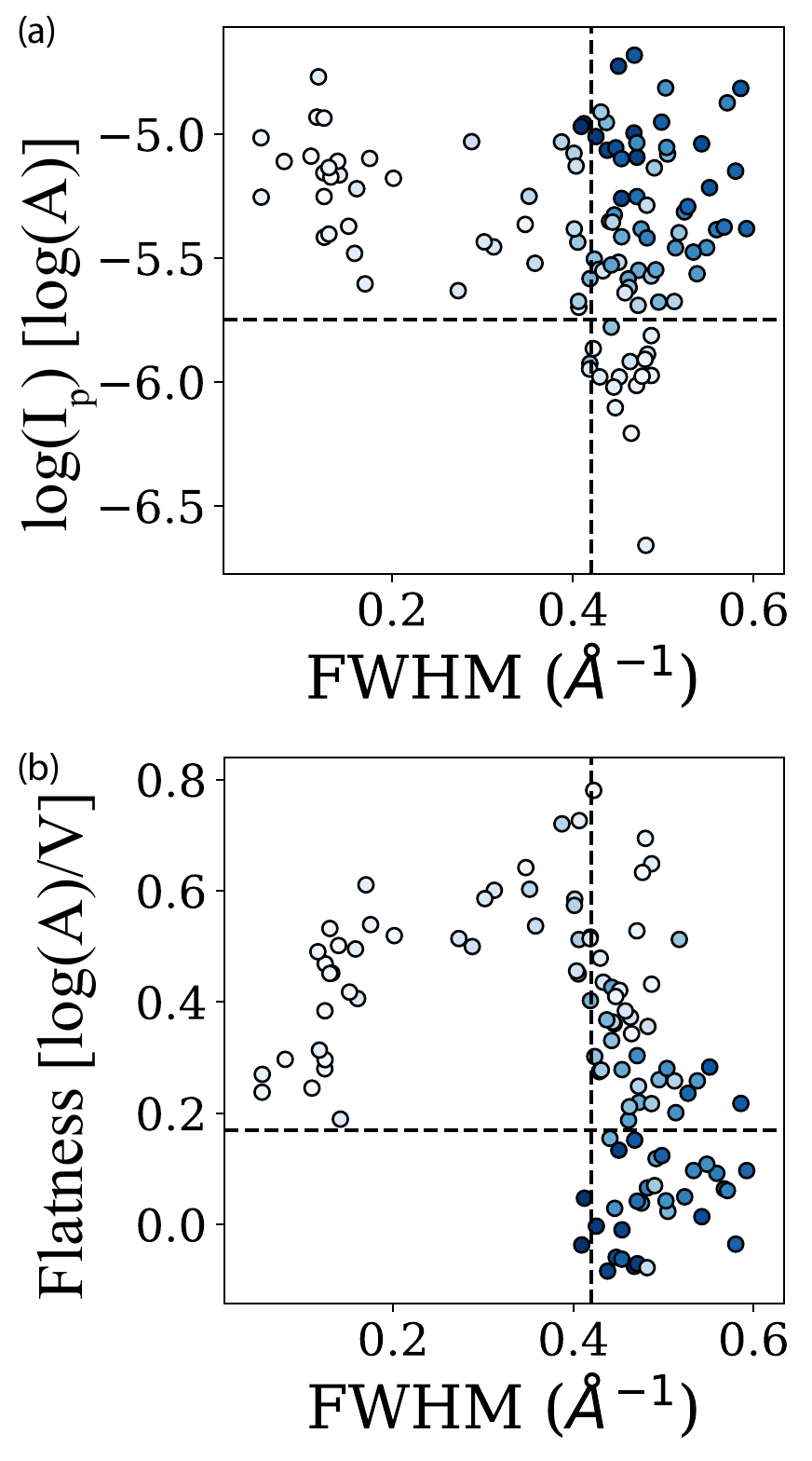}
    \caption{Plots of $\log(I_p)$ and $F_p$ as a function of FWHM of the FSDP. It can be seen that in both cases, the lowest values only exist at a high FWHM.  In (a) $\log(I_p)<-5.75$ and in (b) $F_p<0.17$ require a FWHM$>$0.42 \IA{}. The blue color scale is used to indicated Ti content with darker colors indicating a greater concentration of Ti.}
    \label{fig:correlation}
\end{figure}

Fig. \ref{fig:correlation} plots the $I_{p}$ and $F_{p}$ as a function of FWHM and Ti concentration in the alloy (as color) for the Ni-Ti-Al samples. It is clear in both of these plots that there is a FWHM threshold above which the lowest currents and flatness are observed.  Specifically, for $I_{p}$ we can create a threshold in FWHM at 0.42 \IA{} and in $\log(I_p)$ at -5.75.  Any corrosion current below $\log(I_p)=-5.75$ must have a FWHM above 0.42 \IA{}.  These points can be as much as half an order of magnitude lower than any of the data below the FWHM the threshold.  
Further, there is a clear relationship between the Ti concentration and the corrosion current beyond that threshold. The samples with the lowest Ti concentration (or greatest Al concentration) show the lowest corrosion rates\cite{thomas2003titanium}.  According to the published Pourbaix diagrams, all three elements should form passive oxides at neutral pH in the presence of Cl$^-$ ions\cite{skilbred2016corrosion,davoodi2008multianalytical,bhola2009electrochemical}. In this context, the relative corrosion rates will primarily be determined by the protective nature of the passivating oxide or hydroxide. Similar trends in increased corrosion resistance by the addition of Al have been observed in TiN coatings\cite{li2003electrochemical}.

A similar classification behavior is observed in the $F_p$ vs. FWHM plots, whereby a FWHM of 0.42 \IA{} is necessary, but not sufficient condition, to have a flatness below 0.17 $\log(A)/V$.   Here the opposite compositional variation with Ti/Al concentrations is observed. In this case, the larger Ti concentrations are positively correlated with the smallest flatness values. In this alloy system, there appears to be a trade-off between the corrosion rate (Al concentration) and stability of the oxide to external perturbation (Ti concentration). From the observation of the corrosion properties of the Ni-Ti-Al alloys, and the previous values provided in the literature, it appears that for this system a FWHM of 0.42 \IA is more reasonable for declaring the deposited films as being amorphous. It is interesting to note that a similar trend in corrosion rate and FWHM is observed by \citeauthor{akiyama_habazaki_kawashima_asami_hashimoto_1993}.

\begin{figure}
    \centering
    \includegraphics[width=8.5cm]{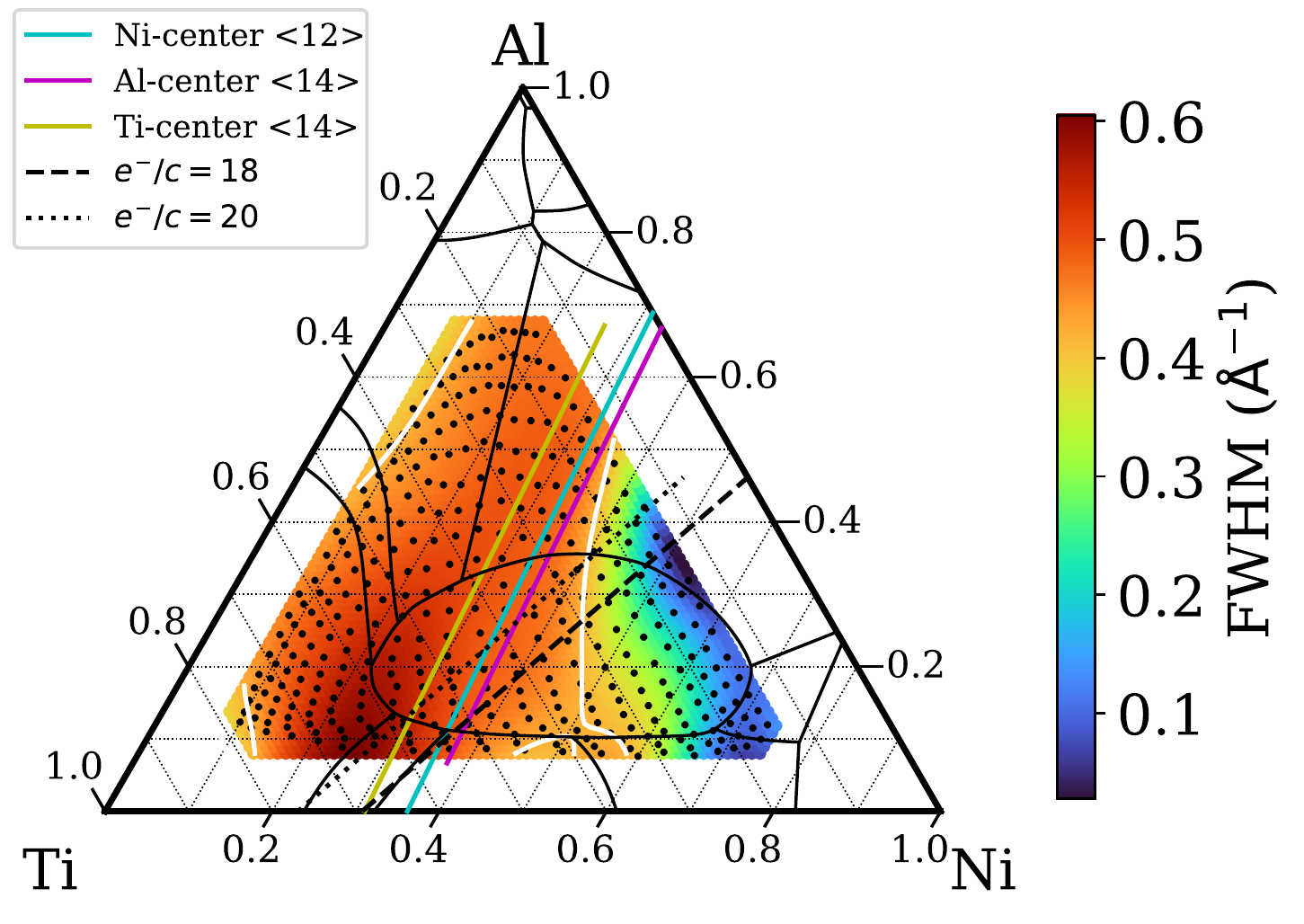}
    \caption{A plot of compositions with efficiently packed clusters and iso-electronic stability based on the model by \citet{laws2015predictive}.  The close packed lines drawn include a Ni-centered cluster with 12 atom coordination, Al-centered cluster with 14 atom coordination, and a Ti-centered cluster with 14 atom coordination.  Isoelectronic lines are drawn showing the number of valence electrons per cluster for the Ni-centered clusters, assuming oxidation states of Ni=0, Ti=2, and Al=3.  The liquidus phase lines from Ref. \citenum{budberg1992aluminium} have also been superimposed onto the phase diagram.}
    \label{fig:Laws}
\end{figure}
Fig. \ref{fig:Laws}, plots the FWHM surface along with lines demarcating the efficiently packed clusters and isoelectronic configurations, and the liquidus projection.  The primary liquidus surface is highly correlated with the liquidus phase field of Ti$_{2}$Ni and indicates that this phase might be difficult to crystallize. The FWHM valley corresponds to a series of phase boundaries and temperature troughs running through the Ti and Al-rich segments of the liquidus surface. This is likely to be associated with topological or electronic stability. Calculations  using the atomic packing efficiency model (Ref. \citenum{laws2015predictive}) were made assuming atomic radii of 125 pm, 142 pm, and 144 pm for Ni, Al, and Ti.  These calculations show an overlap of efficiently packed Ni-centered (icosahedral-like) clusters with a coordination number of $<12>$  with  Al/Ti-centered efficiently packed clusters with a coordination number of $<14>$. There is a distinct correlation between the packing efficiency line, the liquidus lines and the FWHM valley. The valence electrons per cluster, with Ni=0, Ti=2, and Al=3, show very good correlation between e/c=18 and e/c=20 lines and the FWHM. Note that the liquidus lines bounding the Ti$_{2}$Ni phase field follow the two e/c lines. The largest FWHM values reside in the regions that satisfy both the efficient packing and the electrons per cluster models. The FWHM valley follows both the liquidus line and the efficient packing model, possibly indicating a weaker glass forming ability as Al is substituted for Ti.

\section{Summary}

We have performed high-throughput synthesis and structural and electrochemical screening  of a MG Ni-Ti-Al thin-film library.  We show that there is a correlation between the FWHM of the FSDP and the highest corrosion resistance.  We used this correlation to determine that 0.42 \IA{} is a reasonable threshold for FWHM of the FSDP in this system for classifying a material as a MG.  Using this threshold and the FWHM map, we further show that the area we consider to be glassy agrees well with both our ML model as well as the efficient packing model.  While this particular threshold is specific to this system, we believe that this method of using correlation between two screening modalities can be useful in other classification problems based on continuously varying measured quantities.  Our selection of a different criteria for MG classification leads to interesting questions about the  model: How well classified is the cataloged data used to train the ML model?  How much does a change in classification criteria for new high-throughput data affect new predictions?  Most importantly and most generally, how do we handle usage of literature and catalogued data, such as Ref. \citenum{kawazoe1997nonequilibrium}, that do not have the necessary detail to allow for a consistent application of a classification criteria?

A series of bulk samples have been fabricated based on the results from the ML-HTE work described here using vacuum casting.  The molten droplets exhibited high-wetting angle and  slow solidification kinetics, attributes often associated with glass-forming melts. Bulk metallic glass-formation has been confirmed for alloys in the range \ce{Ti_{52.5-62.5}Ni_{30-35}Al_{7.5-15}} of up to 2 mm critical casting diameter.  This work is forthcoming and will be published elsewhere shortly.

\begin{acknowledgement} 
The authors thank Doug van Campen for assistance with collecting the XRD, which was collected at the Stanford Synchrotron Radiation Lightsource, SLAC National Accelerator Laboratory, which is supported by the U.S. Department of Energy, Office of Science, Office of Basic Energy Sciences under contract no. DE-AC02-76SF00515.  H.J. and B.L.D. thank the NIST/NRC postdoctoral fellows program for support.  S.S.  was supported by the Advanced Manufacturing Office of the Department of Energy under grant no. FWP-100250.  The ML work and L.W. were funded under award 70NANB19H005 from U.S. Department of Commerce, National Institute of Standards and Technology as part of the Center for Hierarchical Materials Design (CHiMaD).

\end{acknowledgement}

 \begin{suppinfo}
Data and code used in this work including 1D diffraction data, polarization curves, extracted metrics and data analysis and plotting code can be found here: \url{https://github.com/usnistgov/HTE-NiTiAl}
\end{suppinfo}

\bibliography{HTE_NiTiAl}

\begin{mcitethebibliography}{38}
\bibitem[Not()]{Note-2}
Certain commercial equipment, instruments, software, or materials are
  identified in this paper to foster understanding. Such identification does
  not imply recommendation or endorsement by the National Institute of
  Standards and Technology, nor does it imply that the materials or equipment
  identified are necessarily the best available for the purpose.\relax
\mciteBstWouldAddEndPunctfalse
\mciteSetBstMidEndSepPunct{\mcitedefaultmidpunct}
{}{\mcitedefaultseppunct}\relax
\EndOfBibitem
\end{mcitethebibliography}

\end{document}